# Modeling the Outflow in the Narrow-Line Region of Markarian 573: Biconical Illumination of a Gaseous Disk[1]

T.C. Fischer[2], D.M. Crenshaw[2], S.B. Kraemer[3], H.R. Schmitt[4], M.L. Trippe[5]

## ABSTRACT

We present a study of the outflowing ionized gas in the resolved narrow-line region (NLR) of the Seyfert 2 galaxy Mrk 573, and its interaction with an inner dust/gas disk, based on *Hubble Space Telescope (HST)* WFPC2 and STIS observations. From the spectroscopic and imaging information, we determined the fundamental geometry of the outflow and inner disk, via two modeling programs used to recreate the morphology of these regions imaged with *HST*. We also determined that the bicone of ionizing radiation from the Active Galactic Nucleus (AGN) intersects with the inner disk, illuminating a section of the disk including inner segments of spiral arms, fully seen through structure mapping, which appear to be outflowing and expanding. In addition, we see high velocities at projected distances of $\geq 2''$ ($\sim 700$ pc) from the nucleus, which could be due to rotation or to in situ acceleration of gas off the spiral arms. We find that the true half opening angle of the ionizing bicone (53°) is much larger than the apparent half-opening angle (34°) due to the above geometry, which may apply to a number of other Seyferts as well.

*Subject headings:* galaxies: Seyfert – galaxies: individual (Mrk 573)

[2]Department of Physics and Astronomy, Georgia State University, Astronomy Offices, One Park Place South SE, Suite 700, Atlanta, GA 30303; fischer@chara.gsu.edu

[3]Institute for Astrophysics and Computational Sciences, Department of Physics, The Catholic University of America, Washington, DC 20064

[4]Computational Physics, Inc, Springfield, VA 22151-2110; and Naval Research Laboratory, Washington, DC 20375

[5]Department of Astronomy, University of Maryland, College Park, MD 20742-2421

## 1. Introduction

AGN are believed to be powered by accretion of matter onto supermassive black holes, which occupy the gravitational centers of their host galaxies. Seyfert galaxies, which host relatively low luminosity ($L_{bol} \leq 10^{45}$ erg s$^{-1}$), nearby (z $\leq$ 0.1) AGN, are typically grouped into two classes (Khachikian & Weedman 1974). Seyfert 1 galaxies have spectra containing broad (full width at half-maximum [FWHM] $\geq$ 1000 km s$^{-1}$) permitted lines, narrower (FWHM $\leq$ 1000 km s$^{-1}$) forbidden lines, and distinct, nonstellar optical and UV continua. Seyfert 2s differ in that they contain narrow permitted and forbidden lines, and their optical and UV continua are dominated by the host galaxy, though through spectral polarimetry of Seyfert 2s, broad permitted lines and nonstellar continua have been detected in polarized light (Antonucci & Miller 1985). A unified model for Seyfert galaxies (Antonucci 1993), proposes that Seyfert 1s and 2s differ only by orientation of the AGN with respect to the observer's line of sight. The broad emission-line gas and continuum source are surrounded by a large "torus" of gas and dust, which lies along our line of sight for Seyfert 2s, thus obscuring our view of the broad-line region (BLR) and continuum source in those galaxies. Based on variability, the broad-line region measures several to tens of light days (Peterson et al. 2004), while the narrow-line region, where forbidden lines and narrower components of permitted lines form, can extend out to $\sim$1 kpc (Pogge 1989).

The NLRs of Seyfert galaxies contain gas that has been photoionized by the nonstellar continuum emitted by the AGN and are roughly biconical in structure, with the apex of the bicone residing in the central AGN (Pogge 1988; Schmitt et al. 1994). This suggests that the ionizing radiation is collimated by an absorbing material, which could be the presumed torus or possibly a disk wind (Elitzur & Shlosman 2006), at small radial distances. However, the source of the NLR gas is not well understood. At radial distances of $\geq$1 kpc, the gas in the extended NLR (ENLR) probably lies in the plane of the host galaxy (Unger et al. 1987).

Mrk 573 (UGC 1214) has been thoroughly studied, as it houses a bright AGN. It is a Seyfert 2 galaxy because only narrow (FWHM < 1000 km s$^{-1}$) emission lines are present in its optical spectra. However, spectropolarimetric observations via *Subaru* (Nagao et al. 2004) revealed broad H$\alpha$ (FWHM $\approx$ 3000 km s$^{-1}$), H$\beta$, and Fe II multiples, which confirm the presence of Mrk 573's hidden BLR.

At a redshift of $cz = 5160(\pm10)$ km s$^{-1}$ from H I 21-cm radiation (Springob et al. 2005), Mrk 573 is at a distance of $\sim$72.6 Mpc (for $H_0 = 71$ km s$^{-1}$ Mpc$^{-1}$); at this distance, $1''$ is equal to a tranverse size of 340 pc. The host galaxy is classified as (R)SAB(rs) (de Vaucouleurs et al. 1995) and hosts a triple radio source (Ulvestad & Wilson 1984) which is comprised of a central component and a pair of lobes which lie along a position angle of $\sim$125º. Extended [O III] and [N II] $\lambda\lambda$ 6548, 6583 plus H$\alpha$ emission along the radio axis shows a biconical morphology (Tsvetanov & Walsh 1992; Pogge & De Robertis 1995), similar to a number of other well-studied Seyfert 2 galaxies (Schmitt et al. 2003).

Strong high-ionization forbidden lines ([Fe VII] $\lambda 6087$, [Fe X] $\lambda 6374$, [Ne V] $\lambda\lambda 3346,3426$) have been detected in optical spectra via ground-based observations (Storchi-Bergmann et al. 1996; Erkens et al. 1997), indicating that the NLR is more highly ionized than many Seyfert 2 galaxies (Koski 1978). In fact, high-ionization gas has been found at distances of $\sim$2.5 kpc from the AGN (Storchi-Bergmann et al. 1996). We give further information on the physical conditions in the central emission-line knot of the NLR in Kraemer et al. (2009), henceforth Paper 1.

Ground-based images of Mrk 573 in the V filter (Afanasiev et al. 1996) show a host galaxy disk, with an inclination $i = 30^{\rm o}$ and a major axis position angle $P.A. = 103^{\rm o}$ with the closest edge located in the southwest, based on fits to the gas kinematics at distances $> 6''$ from center. In Figure 1, we show a structure map (containing both continuum and line emission) of an F606W image of Mrk 573, generated by processes described by Pogge & Martini (2002) and Deo et al. (2006), which shows the ionized gas regions of the ENLR to have a morphology of arcs and spiral fragments. These formations appear to be connected to the surrounding dust lanes (Quillen et al. 1999; Pogge & Martini 2002), with the NW section of ionized gas leading to the south spiral arm and the SE section leading to the north arm.

Previous analysis by Schlesinger et al. (2009) used STIS long-slit spectra to map kinematics across the NLR in a study of the outflow and feedback of Mrk 573 and determine that shock ionized radiation is not significant in the disk. Paper I also uses sophisticated photoionization models to show that photoionization dominates as the source of emission for Mrk 573's inner NLR (similar to Mrk 3 (Collins et al. 2005, 2009), NGC 1068 (Kraemer & Crenshaw 2000a,b), and NGC 4151 (Kraemer et al. 2000; Crenshaw & Kraemer 2005, 2007)). In this paper, we use the same data as part of our ongoing study to determine the geometries of the outflows and fueling flows in Seyfert galaxy AGN (Crenshaw et al. 2010; Das et al. 2005, 2006). We chose to study Mrk 573 as its morphology follows the same pattern seen in Mrk 3, another Seyfert galaxy that we have previously studied via STIS long-slit spectra (Crenshaw et al. 2010). With well resolved emission arcs around its nucleus, and the fortunate position of the STIS slit across them, this data set proved to be an excellent opportunity to apply our kinematic and geometrical models.

## 2. Observations

We retrieved archival Space Telescope Imaging Spectrograph (STIS) long-slit spectra and Wide Field and Planetary Camera 2 (WFPC2) images of Mrk 573 from the Multimission Archive at the Space Telescope Science Institute (MAST). Spectra were obtained on 2001 October 17 under Hubble Program ID 9143 (R.Pogge, PI) with a $52'' \times 0.''2$ slit using both a medium-dispersion G750M grating centered on H$\alpha$ emission (6300 - 6850 Å), and a low-dispersion G430L grating (2900 - 5700Å) that includes [O III] $\lambda$5007, the brightest line in the spectrum. The spectral resolutions were 1.1 and 5.5 Å respectively with an angular resolution of $0.''051$ per pixel in the cross dispersion direction. The slit for these spectra has a position angle of $-71.2^{\rm o}$, favorably overlapping the spatially resolved central knot of [O III] and H$\alpha$ emission, similar to knots seen in most nearby Seyfert galaxies (Crenshaw & Kramer 2005), and also crossing several arclike structures in the NLR that have similar strong emission. Figure 1 shows the location of the slit superimposed over the structure map. The location of the optical continuum peak, identified by a $''+''$ in the figure, was found by aligning WFPC2 [O III] and continuum images obtained on the same date (Schmitt et al. 2003). The G750M spectra were taken as a combination of three exposures with times of 1080, 1080, and 840s. The G430L spectra were taken as two exposures of 805 and 840s. Observations were dithered by $\pm 0.''25$ along the slit with respect to the first spectrum to avoid problems due to hot pixels, and wavelength calibration lamp spectra were taken during Earth occultation. A list of all observations, further STIS spectra p rocessing, and cosmic ray removal techniques and resulting calibrated spectral images for both gratings are given in detail in Paper 1.

Using the processed long-slit data, we employed a program to fit the lines with Gaussians over an average continuum taken from line-free regions throughout the spectra (Das et al. 2005). This allowed us to find the central peaks of the Gaussians, which gave us the central wavelengths from which we measured Doppler shifted velocities of both the H$\alpha$ and [O III] emission lines, while also deblending lines that we needed from adjacent, overlapping lines (such as H$\alpha$ deblended from [N II] emission). There were three sources of uncertainty in our velocity measurements, as detailed by Das et al. (2005). The first is that the measured emission lines are not perfect Gaussians, the second comes from emission cloud displacements from the center of the $0.''2$ slit in the dispersion direction, and the third comes from noisy spectra. Errors were converted to velocities and added in quadrature, which produced a total maximum error of $\pm$ 60 km s$^{-1}$. Extremely noisy spectra (spectra without detectable emission $> 3\sigma$) between $0.''75$ and $1.''75$ on either side of the nucleus, as well as $\geq 4.''75$ to the northwest and $\geq 3.''5$ to the southeast of the nucleus, were not fitted. By calculating velocities from line centroid shifts and subtracting the systemic velocity of the galaxy, we were able to determine the radial velocities along the slit in the rest frame of Mrk 573.

## 3. Results

Figure 2 shows the rest frame radial velocities, FWHM, and normalized fluxes along the slit for both [O III] $\lambda5007$ and H$\alpha$ $\lambda6563$ emission lines. The three dashed lines represent the central positions of the emission-line arcs from flux intensity. Discrepancies between the FWHMs of [O III] and H$\alpha$ are due to different resolutions of the gratings. Adding the FWHM of the resolved H$\alpha$ lines ($\sim 250$ km s$^{-1}$) to the FWHM of the line spread function of the G430L grating for a $0\rlap{.}''2$ slit ($\sim 650$ km s$^{-1}$) in quadrature gives a value typical of the observed FWHM of the [O III] lines ($\sim 700$ km s$^{-1}$). Comparisons with work by Schlesinger et al. (2009) show that our measurements are essentially in agreement. Several interesting features are present in both spectra. On either side of the nucleus are high radial velocities that show distinct asymetrical red/blue shifts that are characteristic of biconical outflows seen in the Seyfert 2 galaxies NGC 1068 (Das et al. 2006) and Mrk 3 (Ruiz et al. 2001; Crenshaw et al. 2010). Between $0\rlap{.}''75$ and $1\rlap{.}''75$ from the nucleus there is an absence of substantial emission gas (see the [O III] image of Paper 1); thus we have no measurements in that span. Beyond the empty region, there is emission between 2-4$''$ on either side of the nucleus that produces linear velocity curves with amplitudes reaching 200 km s$^{-1}$ that Schlesinger et al. (2009) have suggested are due to emission in a rotating disk.

## 4. Models

We assume the ionizing radiation responsible for the NLR and ENLR is a bicone (where we define half of the bicone as an individual cone), because it is the simplest geometric shape produced by a central obscuring torus. As the STIS slit passes over the NLR emission, we see kinematic components for both sides of the bicone, a front and back to each half, indicating that it has a hollow core. We used our kinematics modeling program from Das et al. (2005), which allows us to recreate the observed radial velocities along a fixed slit position by altering various physical parameters of our model bicone outflow. Our model included 7 alterable parameters, initial values came from the [O III] image from Paper 1 (deprojected height of bicone [$z_{max}$], outer opening angle [$\theta_{max}$], and position angle) and kinematics from Figure 2 (maximum velocity [$v_{max}$] and turnover radius [$r_t$]), allowing inner opening angle ($\theta_{min}$) and inclination of the bicone axis out of the plane of the sky to be free parameters. We adopted our previous velocity law, a linear increase and subsequent linear decrease with distance, because it is the simplest law that matches the observations.

We started with the narrow bicone observed in the [O III] image of Paper 1 ($\theta_{max} = 68°$), which proved to be difficult to model as we adopted an acceleration / deceleration model which fit the inner, high radial velocity curve at $< 1''$ as accurately as possible. In doing so, the outflow of each half of the bicone was either entirely redshifted or blueshifted. This was an unacceptable model as Figure 2 shows kinematics on either side of the nucleus that are both red and blueshifted. Thus, in order to match our model with the data, we needed to widen the opening angle such that the kinematic component nearest the plane of the sky for each bicone half passed beyond the plane so that its outflow was shifted in the opposite direction of the inner, high velocity kinematic component. To widen the opening angle while maintaining the accurate fit of this inner component required an additional adjustment of the inclination away from our line of sight so that the modeled outflow kinematics near the nucleus would remain successfully fit. Furthermore, while attempting to fit our kinematics model to our data, we needed to account for the circumstances of Mrk 573 uncharacteristic of normal bicone outflow. Preferably, kinematics of the outflow contain four peaked radial velocity curves (Das et al. 2005, 2006), whereas only the two inner, high velocity curves in the kinematics of Mrk 573 contain peaks. In the outer, lower velocity curves, we see a linear increase in radial velocities which our modeling program was not able to account for. Fits for these outer velocity curves were made such that the increasing slope of the model coincided with the data set, as shown in our model that best fits the data in Figure 3.

After creating a reasonable kinematics model, the actual outer opening angle of the extended emission at distances $> 1.''$ was found to be much larger than the narrow opening angle we observe in Figure 1 and the [O III] image of Paper 1. From the presence of arcs in the ENLR that appear to extend into the outer dust lanes of the galaxy, and similar circumstances occuring in Mrk 3 (Crenshaw et al. 2010), it is likely that the opening angle seen in images of Mrk 573 is a projection effect due to an intersection of the outflow and the inner galactic disk. This intersection provided us with an additional constraint in which to define the geometry of the emission line regions. We used the best-fit model bicone's $\theta_{max}$, position angle, and inclination values to create a simple, three-dimensional geometric model of the ionizing bicone's outer surface. We then included a two-dimensional galactic disk, using the observed inclination and postion angles, 30º and 103º respectively (Afanasiev et al. 1996; Kinney et al. 2000), in the model to show the likely spatial interaction between the two components. This created a projected opening angle upon the model disk that we compared with the emission seen in both the structure map in Figure 1 and the [O III] WFPC2 image of Paper 1. Establishing this second model allowed us to check whether our kinematics model then provided an accurate description of emission seen in imaging of Mrk 573. We altered our parameters in an iterative process until a reasonable kinematic model corresponded with a geometric model whose bicone edges were within 5º of the imaged emission region, a discrepency that could be due to the unmodeled thickness of the disk. The model that fits this requirement is shown in Figure 4 and a frame of the model is superimposed over the structure map from Figure 1 in Figure 5 to show how well it fits over the imaged emission, where shaded regions represent portions of the bicone not covered by the disk. Though our model uses a two-dimensional plane for the galactic disk, we can assume a typical dust scale height value of ∼200 pc for spiral galaxies (Xilouris et al. 1999) and a length of ∼1500 pc for the bicone along its axis taken from the [O III] image of Paper 1, allowing for much of the bicone edge that lies near the plane of the sky to still interact with the disk. Thus, a reasonable thickness for the inner galactic disk could explain the small discrepancies between our model and the observations in Figure 5.

Final values of all model parameters, as well as the observed inclination and position angle of the disk as given by Afanasiev et al. (1996) are provided in Table 1. Clearly, the geometrics and kinematic constraints of our models lead to significantly different parameters for the ionizing bicone compared to the apparent values (PA = −36º vs. −52.5º, $\theta_{max}$ = 53º vs. 34º).Though our models are in fair agreement on the parameters which define Mrk 573's bicone, additional STIS spectra at another position angle near -60º would allow us to refine the model and compare it against the new set of velocities to check for an accurate match. A simple check would be to look for increased peak velocities for a slit near the axis of the model bicone (PA = −36º), where we would see the maximum possible velocities from our point of view.

## 5. Discussion

We have shown how our kinematics and geometric models of Mrk 573 determine the orientation of its biconical outflow with respect to our line of sight and the galactic disk. The remaining issue in our kinematics model is the lack of deceleration in the outer, low velocity kinematic components, at projected distances > 2″ (∼700 pc), as shown in Figure 3. Where the bicone kinematics of our previously studied AGN contain four sets of accelerating and decelerating outflows (Das et al. 2005, 2006; Crenshaw et al. 2010), only Mrk 573's two inner kinematic components follow this pattern. We suggest that the easily fit kinematics components ±1″ are dominated by emission from the outflow, while extended emission outside ±1″.75 is dominated by disk gas interacting with the bicone of ionizing radiation. As such, it is the presence of the disk that may be responsible for the absence of peaks present in the outflows of other Seyfert NLRs.

One possible reason we see large radial velocities at distances > 2″ is that they are due to emission from a rotating disk, which our modeling program does not account for (Schlesinger et al. 2009). This suggestion immediately seems plausible as images like Figure 1 show emission from arcs within the ENLR that may be joined with dust lanes further out. However, there are a couple of potential problems with this hypothesis. With Mrk 573's inclination being 30º out of the plane of the sky, a deprojection of the extended velocities would reach speeds of ∼400 km s$^{-1}$. Sparke & Gallagher (2000) note that while velocities this large are not unheard of, with the fastest measured rotation being ∼ 500 km s$^{-1}$ in UGC 12591, they are rare. Using a warped disk (Lawrence & Elvis 2010) in an attempt to explain the kinematics via rotation could bring down the deprojected velocities. Warping the inner disk further out of the plane of the sky would reduce the difference between the projected and actual velocities creating a more plausible argument for the outer kinematics to be due to rotation. What warping does not fix, however, is the direction in which the disk is rotating. Our kinematics data show the southeast end of the slit to be blueshifted and the northwest end to be redshifted. However, Mrk 573's clockwise spinning morphology in Figure 1, along with having the southwest side of the disk being nearest to us, suggests that the northwest region should be blueshifted and southeast redshifted, as nearly all spiral galaxies rotate in the sense that their arms appear to "wind up" (de Vaucouleurs 1958; Toomre 1981). Thus, Mrk 573 may be a candidate for showing leading spiral structure which, similar to NGC 4622 (Buta et al. 2003), may result from a minor merger or tidal encounter. However, once again, this would be a rare circumstance. We also checked the possibility that the disk is inclined in the opposite direction (−30º instead of 30º). Making this alteration in the geometrical model has the disk almost parallel with the bicone axis, effectively bisecting each half. This is a problem because we do not see a projected opening angle large enough to agree with the model (roughly twice as wide). Furthermore, this model does not agree with our finding from the kinematics data that the bicone is hollow along its axis Both problems led us to conclude that a reversed inclination is not feasable. H I 21-cm observations would likely be the best means to clarify the large scale rotational structure.

If these kinematics are indeed not due to rotation, another option is in situ acceleration of gas off the emission-line arcs due to radiative driving or entrainment by highly ionized winds. Work by Das et al. (2007) on NGC 1068 shows radiative acceleration can drive ionized clouds of gas outward at these large distances. Although we do not have any kinematic data between 0.″75 and 1.″75 from the nucleus, there are three velocity gradients (centered on dashed lines located at ∼ −2.″75, −2″, and 1″ in Figure 2) that coincide with the positions of the emission-line arcs along the slit seen in the structure map of Figure 1. All three arcs' velocities are redshifted on the southeast side and blueshifted on the northwest side. These gradients are possibly due to the superposition of the slit on a projected outward expansion of the arcs occuring along the intersection between the arcs and the edge of the bicone flowing through the disk. This interaction would create curved annluar shells around the edge of the bicone that, from our point of view and using the given position of the slit, could allow us to see the expanding shell's rear redshifted region first and then the front blueshifted region as we go from the southeast end of the long-slit to the northwest. This is strong evidence that the emission-line gas may be accelerating off the arcs, which are likely continuations of the dust spirals and may represent the original fueling flow to the AGN, as we have suggested for Mrk 3 (Crenshaw et al. 2010).

Comparing our opening angle against the projected opening seen in *HST* images, as well as against opening angles given in previous works (Schlesinger et al. 2009; Wilson & Tsvetanov 1994), our value is much larger and consistent with those of other recently studied Seyfert 2s, Mrk 3 and NGC 4151 (Das et al. 2005; Crenshaw et al. 2010). These examples may not be unique and large opening angles are likely common for Seyfert galaxies. In targets where most of the observed NLR / ENLR emission comes from the interaction between the ionizing radiation and the galactic disk, similar to Mrk 573 and Mrk 3, observed opening angles can be much smaller than their true opening angles. Data from Mrk 573, Mrk 3, NGC 4151, and NGC 1068 (Das et al. 2006) support the fact that we only see [O III] and Hα around the edges of the bicone. Thus, if observed targets have a biconical outflow that is not apparently hollow at the center (ie a filled triangular projection of emission in the plane of the sky) and show evidence of an interaction with the galactic disk, we can assume an intersection between the galactic disk and an edge of the bicone. Intersections closer to the axis of the bicone would result in projecting an illuminated "V" as the disk interacts with the edges of the outflow bicone and not the inner, hollowed region. This being said, any observed opening angle for targets matching the above description, similar to Mrk 573, would likely be smaller than the true opening angle.

Osterbrock & Shaw (1988) used statistics of numbers of Seyfert 1, 1.5, 1.8, 1.9, and 2 galaxies in Wasilewski field to find that 78% of field AGN were type 2, leading them to calculate an average torus half-opening angle of approximately 35º. With Mrk 573, we now have three very clear type 2 Seyferts objects with opening angles bigger than this calculated average opening angle (NGC 1068:40º, Mrk 3:51º, and Mrk 573:53º). Schmitt et al. (2001) calculated a half opening angle of 48º, a value similar to our results, by binning Seyfert 1.8s and 1.9s with Seyfert 2s to determine the ratio of Seyfert 1s to 2s. Continued research on opening angles of individual Seyfert galaxies is required to determine if these large opening angles are unique. Should we find values consistently above those determined from the Seyfert 1 / Seyfert 2 ratio, we must question if simple obscuration by tori explains the relative fraction of Seyfert 1s to Seyfert 2s. For the opening angle to be so large and depend solely on the torus for its definition, the estimated percentage of Seyfert 2s observed would be less than what we actually see. It is possible that some secondary property could be accountable for the fraction of Seyfert 2s that we see, such as the presence of obscuring gas which is separate from the torus, although possibly connected to the creation of dusty circumnuclear gas that makes a torus.

There is some uncertainty on whether luminosity affects the opening angle of the torus. Investigation of luminosity dependence, such as the "receding torus" model (Lawrence 1991), versus luminosity independence (Grimes et al. 2004; Lawrence & Elvis 2010) debates would be greatly aided by a survey of the bicone opening angle / luminosity relationship of various AGN. As studies of Mrk 3, NGC 1068, NGC 4151, and Mrk 573 all result in outflow bicones of near the same opening angle and house AGN of similar luminosities, further research into less luminous Seyferts and more luminous quasars would allow us to see if there is any clear dependence between bicone (and thus torus) opening angle and luminosity.

## 6. Conclusions

We have analyzed STIS long-slit G430L and G750M spectra of the inner emission knots and nucleus of the Seyfert 2 galaxy Mrk 573. We generated kinematics and geometrical models of the NLR and ENLR, and were able to fit them successfully to our data. Along with determining parameters of Mrk 573's outflow bicone, given in Table 1, we have found that most of the circumnuclear emission comes from the intersection of the galactic disk with the bicone of ionizing radiation. This is supported by the presence of emission arcs that appear to be related to the outer dust lanes as well as the success of our geometrical model. Similarities between Mrk 573 and Mrk 3 suggest that the emission could be due to ionization of the original fueling flow (Crenshaw et al. 2010).

Kinematics outside projected distances of 700 pc from the nucleus could possibly be due to rotation, although this seems unlikely due to a large deprojected amplitude, large velocity dispersion, and requirement of leading spiral structure. Alternatively, we might be seeing in situ acceleration of gas off the previously non-emitting inner dust/gas spiral arms. The latter more easily explains the large velocity dispersions at each emission arc.

The half-opening angle we find from our models (53°) is similar to previously studied Seyfert 2s Mrk 3 and NGC 4151 and is larger than the ratio of Seyfert 1s to Seyfert 2s would predict. Additional modeling of the interaction between the NLR outflow and the galactic disk supports our suggestion that any observed opening angle for Seyfert galaxies exhibiting an interaction between the outflow and the galactic disk would likely be much smaller than the true value due to projection effects. If continued research on determining geometrical parameters in Seyfert galaxies finds opening angles greater than both observed and statistically predicted, obscuration from non-toroidal components might be accountable for the large percentage of Seyfert 2 galaxies that we see.

Table 1.   Best Fit Model Parameters for Mrk 573[a]

| Parameter | Values |
|---|---|
| Disk | |
| $P.A.$ | 103º |
| $i$ | 30º(SW) |
| Bicone | |
| $P.A.$ | −36º |
| $i$ | 30º(NE) |
| $\theta_{max}$ | 53º |
| $\theta_{min}$ | 51º |
| $v_{max}$ | 400 km s$^{-1}$ |
| $z_{max}$ | 1200 pc |
| $r_t$ | 800 pc |

[a]The letters in parentheses indicates the side closest to us.

Fig. 1.— Enhanced contrast $20\rlap{.}''\times20\rlap{.}''$ structure map of the *HST* WFPC2 image of Mrk573, obtained with the F606W filter. Bright areas correspond to line emission and dark areas correspond to dust absorption. Solid lines outline the position of the STIS slit, dashed line depicts the P.A. of the radio components.

Fig. 2.— [O III] and Hα plots showing radial velocities (top), FWHM (middle), and normalized total flux (bottom). Hashed lines depict the position of the two southeast arcs, the nucleus, and the single northwest arc. Note the broader features of the [O III] G430L grating due to the poorer resolution in low dispersion.

Fig. 3.— Kinematics model chosen as the best fit for our radial velocity data set. Parameters used to create this model are given in Table 1. Extended, lower velocity measurements outside $2\rlap{.}''$ from the nucleus are not enclosed in the model parameters as our model does not account for interactions with a disk

Fig. 4.— Geometric model of the NLR and inner disk in Mrk 573, based on parameters from Table 1, shown as viewed from Earth.

Fig. 5.— Structure map of Mrk 573 with superimposed geometrical model. Shaded regions represent the surface of the bicone in front of the disk intersection along our line of sight.

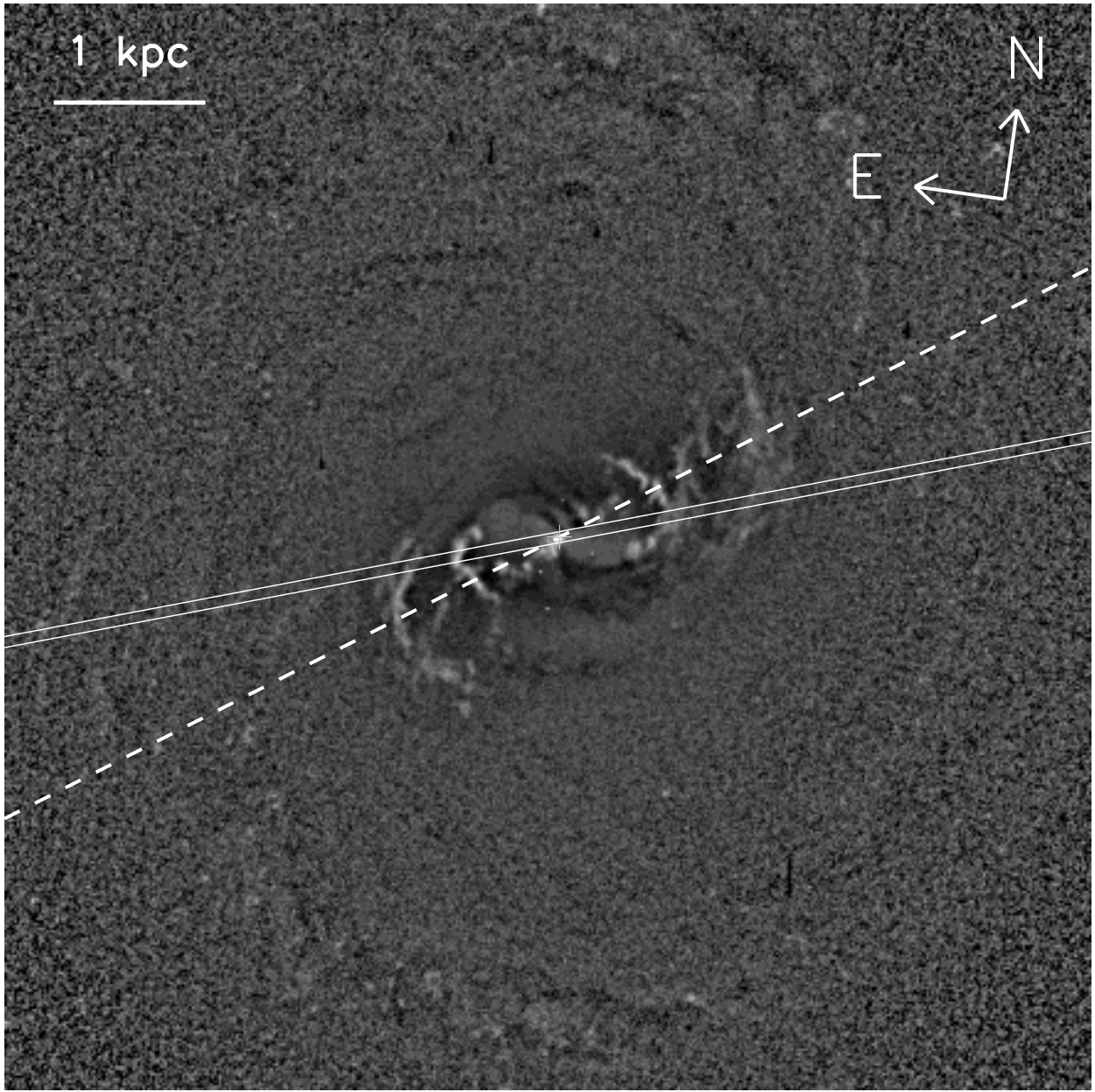

Fig. 1.

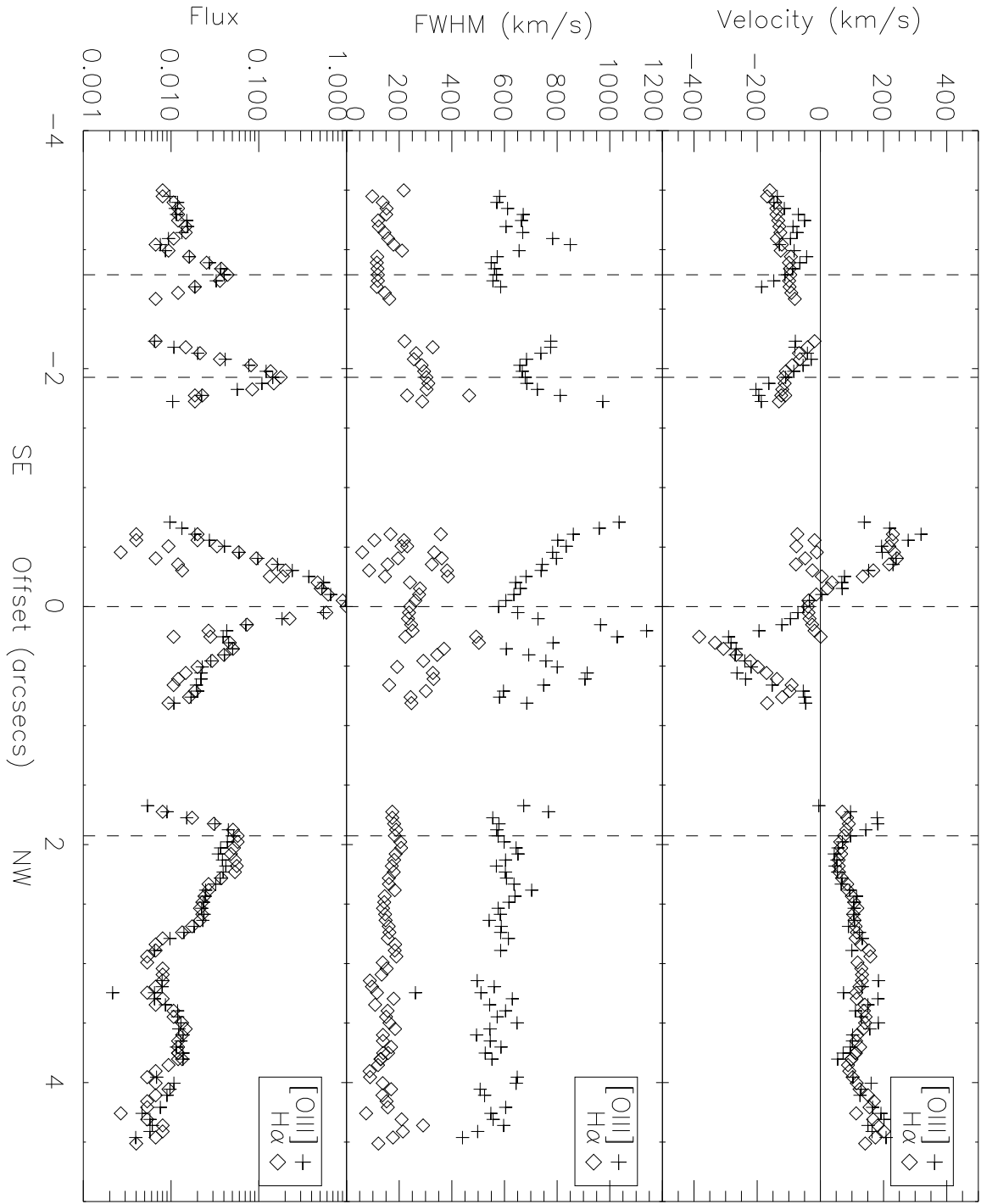

Fig. 2.

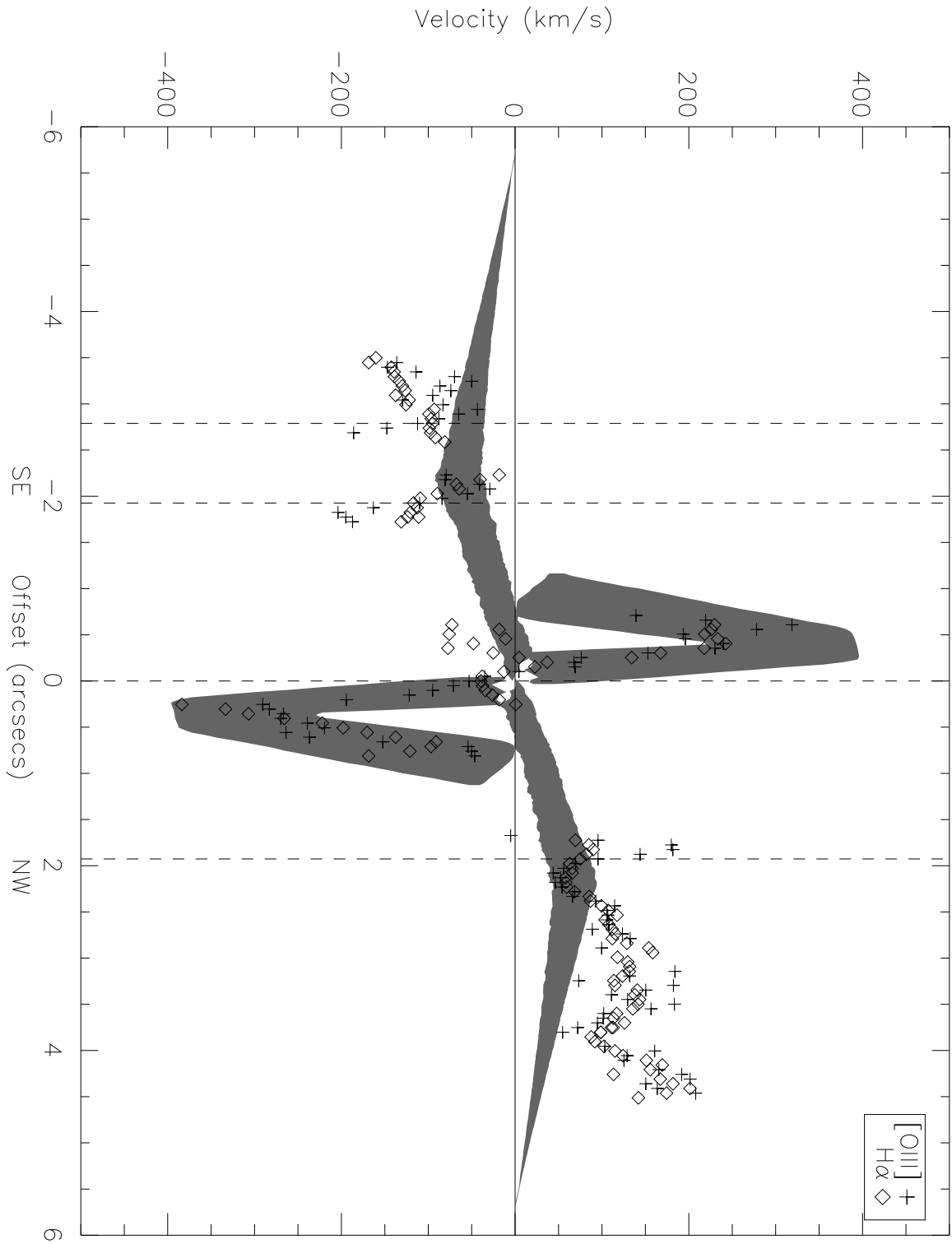

Fig. 3.

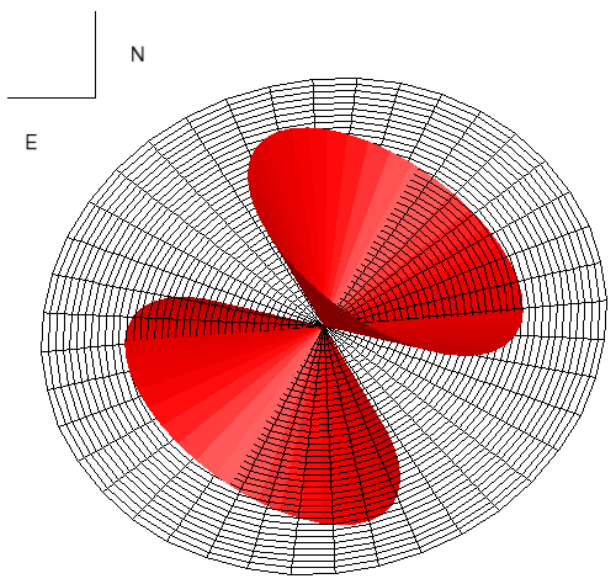

Fig. 4.

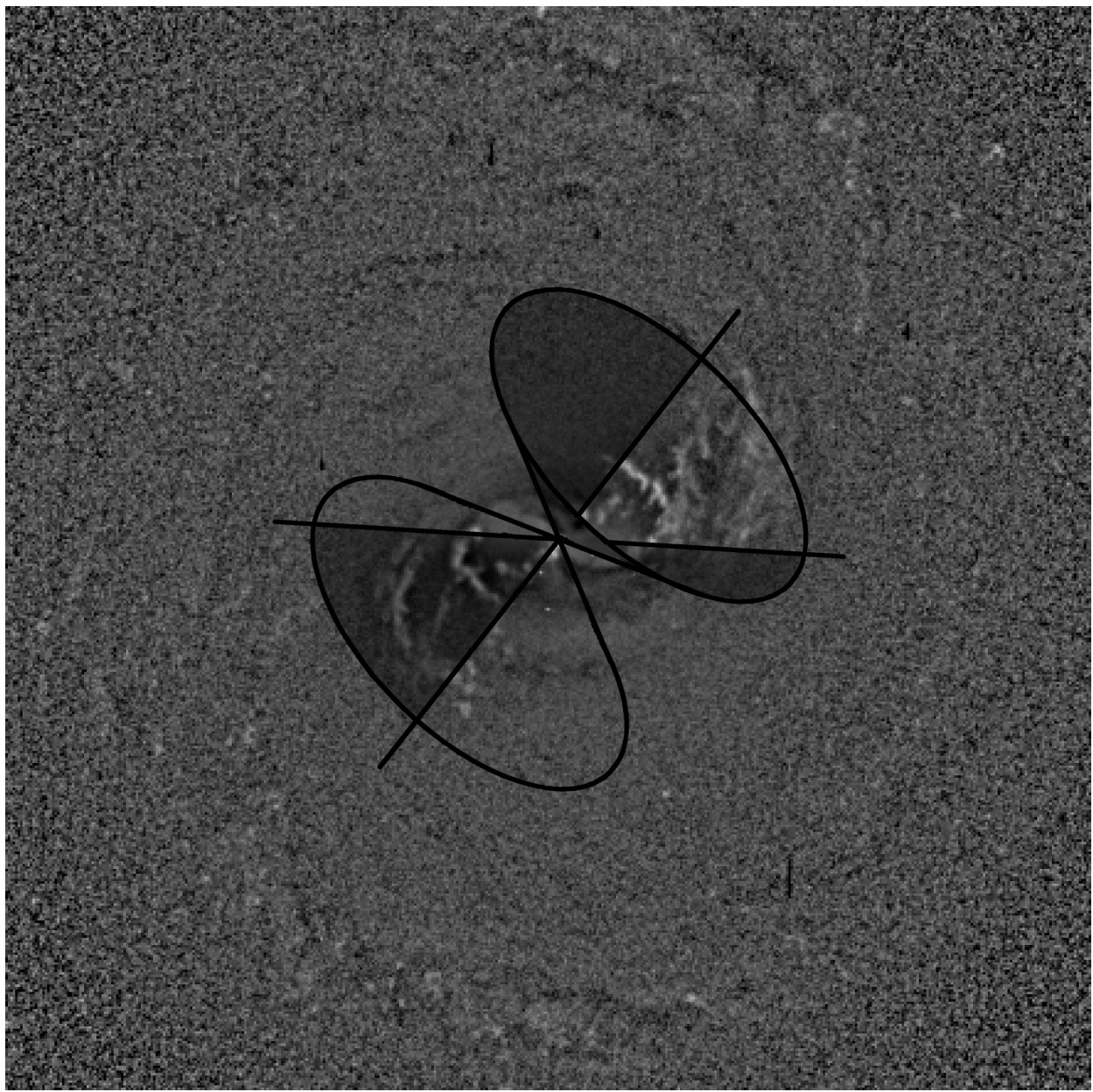

Fig. 5.